\documentclass[prb]{revtex4}

\usepackage{graphicx}
\usepackage{dcolumn}
\usepackage{bm}

\begin{document}

\preprint{APS/123-QED}

\title{Anisotropy in the Incommensurate Spin Fluctuations of Sr$_2$RuO$_4$ 
}

\author{T. Nagata}
\affiliation{%
Department of Physics, Ochanomizu Univ., Tokyo 112-8610, Japan}%

\author{M. Urata}
\affiliation{%
Graduate School of Humanities and Sciences, Ochanomizu Univ., Tokyo 112-8610, Japan}%

\author{H. Kawano-Furukawa}
\affiliation{%
Department of Physics, Ochanomizu Univ., Tokyo 112-8610, Japan}%
\affiliation{%
PRESTO, Japan Science and Technology Corporation, Kawaguchi, Saitama 332-0012, Japan}%

\author{H. Yoshizawa}
\affiliation{%
Neutron Science Laboratory, I.S.S.P., The University of Tokyo, Ibaraki 319-1106, Japan}%

\author{H. Kadowaki}
\affiliation{%
Department of Physics, Tokyo Metropolitan University, Tokyo, 192-0397, Japan}%

\author{P. Dai}
\affiliation{%
Department of Physics and Astronomy, The University of Tennessee, Knoxville, Tennessee 37996, USA}%

\date{tentative version \today}

\begin{abstract}

It has been proposed that Sr$_2$RuO$_4$ exhibits spin triplet superconductivity mediated by ferromagnetic fluctuations. So far neutron scattering experiments  have failed to detect any clear evidence of ferromagnetic spin fluctuations but, instead, this type of experiments has been successful in confirming the existence of  incommensurate spin fluctuations near $\textbf{q}=(\frac{1}{3} \ \frac{1 }{3} \ 0)$. For this reason there have been many efforts to associate the  contributions of such incommensurate fluctuations to the mechanism of its superconductivity. Our unpolarized inelastic neutron scattering measurements revealed that these incommensurate spin fluctuations possess  $c$-axis anisotropy with an anisotropic factor $\chi''_{c}/\chi''_{a,b}$ of $\sim 2.8$. This result is consistent with some theoretical ideas that the incommensurate spin fluctuations with a $c$-axis anisotropy can be a origin of $p$-wave superconductivity of this material.
\end{abstract}

\maketitle

\section{Introduction}

Sr$_{2}$RuO$_{4}$  is the first 2D perovskite oxide material known to exhibit  a superconducting transition without containing copper \cite{Maeno94}. While Sr$_{2}$RuO$_{4}$  is isostructural with the high-$T_{c}$ material La$_{2-x}$Sr$_{x}$CuO$_{4}$, its normal state  shows Fermi liquid behavior and it's  superconducting state is not a spin singlet  ($S$ = 0) as observed in the conventional $s$-wave ($l$ = 0) superconductors  or the $d$-wave ($l$ = 2)high-$T_{c}$ materials. Its superconducting state is instead a spin triplet  ($S$ = 1)  with (most-likely) $p$-wave symmetry ($l$ = 1) (see Ref. \cite{review2003} for a recent review).
 
 $\mu $SR and NMR (Knight shift) measurements  have provided experimental evidence of the spin triplet pairing in Sr$_{2}$RuO$_{4}$. $\mu $SR measurements succeeded in confirming the existence of the spontaneous magnetic field below the superconducting transition temperature $T_{c} \sim$ 1.5 K, indicating the time-reversal symmetry-breaking in superconducting state \cite{Luke98}. Knight shift measurements for the oxygen site in the RuO$_{2}$ planes revealed that the spin susceptibility remains temperature independent even below $T_{c}$ \cite{Ishida98}.

Since $T_{c}$ of Sr$_{2}$RuO$_{4}$ ( $\sim$ 1.5 K) is strongly suppressed by nonmagnetic impurities \cite{nonmagimpurity},a mechanism other than electron-phonon interaction was proposed as the origin of the pairing mechanism of the superconductivity observed in this system.   From the analogy with the super-fluid state of $^3$He and from the fact that the neighbor material SrRuO$_{3}$ is  ferromagnetic, it was speculated that Sr$_{2}$RuO$_{4}$ exhibits spin triplet superconductivity mediated by ferromagnetic fluctuations. Up to this date, however, there is no clear experimental evidence of ferromagnetic fluctuations in this  material \cite{Sidis99,Servant01,Braden02}.

The electronic structure of Sr$_{2}$RuO$_{4}$ ($d$-electron system) is much simpler than those of other spin triplet superconductors \cite{Upt3,Uni2Al3}. This fact has stimulated theoretical efforts on the topics of spin triplet superconductivity and the symmetry of superconducting order parameters.

Mazin and Singh have calculated the electronic band structure of Sr$_{2}$RuO$_{4}$ based on the $t_{2g}$ orbital of the Ru$^{4+}$ ($4d^{4}$) and showed that the Fermi surfaces  consist of quasi one-dimensional $\alpha $, $\beta $ planes defined up by the $d_{yz}$, $d_{zx}$ orbital, and of two-dimensional $\gamma$ planes defined the $d_{xy}$ orbital \cite{Mazin97}.  These predictions are consistent with the results of dHvA experiments \cite{Mackenzie96}.  Furthermore, the theory  predicted that  sizable nesting effects in the quasi one-dimensional bands ($\alpha $, $\beta $ planes) may cause the enhancement of the spin susceptibility  near the incommensurate  propagating vector $\textbf{q}$ = $(\frac{1}{3} \ \frac{1}{3} \ 0)$ \cite{Mazin99}. 
Such an enhancement was indeed confirmed in dynamical spin susceptibility $\chi''$($\textbf{q}_{0}, \omega$) at $\textbf{q}_{0}$ = (0.3 0.3 0)  by inelastic neutron scattering (INS) experiments \cite{Sidis99}. These results stimulated discussions about the possibility of $p$-wave superconductivity mediated by such incommensurate spin fluctuations. 

Some theoretical works reported that, if such an incommensurate spin fluctuations possess  $c$-axis anisotropy, the spin triplet superconductivity could be stabilized by such fluctuations \cite{Kuwabara00,Sato00,Kuroki01}. It is therefore of great importance to stablish if there is any observable anisotropy in $\chi''$($\textbf{q}_{0}$) that can be related to the origin of the spin triplet superconductivity in Sr$_{2}$RuO$_{4}$.

Ishida \textit{et al.} have reported the observation of the anisotropic behavior of the spin susceptibility measured by the NMR technique \cite{IshidaYY}. In NMR measurements, one can observe the $\textbf{q}$-integrated spin susceptibility, $\sum_{\textbf{q}} \frac{\chi "(\textbf{q},\omega)}{\omega}|_{\omega\to 0} $. 
Judging from the similarities with the INS data reported by Sidis \textit{et al.} \cite{Sidis99}, Ishida et al attributed the  temperature dependent part of the $\textbf{q}$-integrated spin susceptibilities (observed by the NMR) to the spin susceptibility at $\textbf{q}_{0}$, and reported that $\chi"(\textbf{q}_0,\omega)$ has a $c$-axis anisotoropy with an anisotropic factor $\chi"_c / \chi"_{a,b}$, ( $\chi "_{IC,out}(\textbf{q}_{0},\omega) / \chi "_{IC,in}(\textbf{q}_{0},\omega)$ in their notation) of $\sim 3$.

In order to ascertain the anisotropic nature of the incommensurate spin fluctuations in Sr$_2$RuO$_4$, however, it is necessary to measure its $\textbf{q}$ dependent spin susceptibility $\chi''$($\textbf{q}$, $\omega$) using INS. 
We have performed such measurement and found that the dynamical spin susceptibility of this system at $\textbf{q}_{0}$ indeed exhibits $c$-axis anisotropy with an anisotropic factor of $\sim$ 2.8. This value is consistent with the anisotropic factor estimated from the NMR measurements \cite{IshidaYY}.  Our conclusion is different from those of the recent reports by  Servant \textit{et al.} and  Braden \textit{et al.} \cite{Servant01,Braden02}, this will be discussed at the end of this paper.

\section{Experiment}

\subsection{Sample preparation and Experimental setup}
For this neutron scattering study, we grew large single crystals  of Sr$_2$RuO$_4$ by the floating zone method. The crystals were cut into smaller cylindrical pieces (4 mm$\phi$ in diameter and 30 mm in length). We performed resistivity measurements on these crystals using a Quantum Design PPMS  instrument equipped with a $^3$He option.  These measurements revealed that $T_{c}$(onset) of all the samples lies between  1.4 $ \sim$ 1.6 K. 

The unpolarized INS experiments were performed using the triple axis spectrometer GPTAS installed at the JRR-3M reactor at the Japan Atomic Energy Research Institute (JAERI) in Tokai, Japan. Neutrons with a fixed final momentum of $k_{f}$ = 3.83 \AA$^{-1}$ and a combination of horizontal collimations of 40'-80'-40'-80' (FWHM from the monochromator to the detector) were utilized. A pyrolytic graphite (PG) filter was placed after the sample position to eliminate higher order wavelength contaminations. Three  sets of crystals were prepared in order to probe three different scattering planes, ($h$ $k$ 0), ($h$ $h$ $l$) and (0.7$h$ 0.3$h$ $l$). The total volume of each of these sets was $\sim$ 3 cm$^{3}$ \cite{comments02}. The crystals were sealed in aluminum cans (filled with He exchange gas to ensure a uniform temperature) that were attached to the cold head of a closed-cycle He gas refrigerator. Throughout this paper the scattering vector $\textbf{Q}$ = ($Q_h$ $Q_k$ $Q_l$) is indexed in reduced lattice units with tetragonal reciprocal lattice of $a^{*}$ = $b^{*}$ = 1.63 \AA$^{-1}$ and $c^{*}$ = 0.49 \AA$^{-1}$, respectively.

\subsection{Magnetic neutron scattering}

In this section we describe the method that we used to measure the anisotropic factor of the spin susceptibility $\chi''(q,\omega)$. In a magnetic neutron scattering experiment \cite{Book},  the scattering intensity $I$ is given by 
\begin{equation}
I \propto f_{Q}^{2} \times G,
\end{equation}
where $f_{Q}$ is the magnetic form factor, which is the $Q$ component of the fourier transform of the distribution of unpaired electrons that contribute to the magnetism in the system. If the electronic distribution is isotropic, $f_{Q}$ shows a monotonic decrease with the absolute value of the  scattering vector $\textbf{Q}$, $Q$ as demonstrated in Fig. 1(a). The quantity G in Eq.(1) is an orientation factor related to the fact that neutrons are scattered only by  the magnetic components perpendicular to the scattering vector $\textbf{Q}$.  In the present study, we assumed that the spin susceptibility within the RuO$_{2}$ planes in tetragonal  Sr$_2$RuO$_4$ is isotropic ($\chi''_{a}=\chi''_{b}=\chi''_{a,b}$).  The orientation factor G is then given by 
\begin{equation}
G(\theta) = (1 + \mathrm{sin}^{2} \theta)\ \chi''_{a,b}(\textbf{q},w) + \mathrm{cos}^{2}\theta\ \chi''_{c}(\textbf{q},w),
\end{equation}
where $\theta$  is  the angle between the  scattering vector $\textbf{Q}$ and the $ab$ plane, which changes through the $Q_{l}$ component of the scattering vector. 

In Fig. 1(b)  we show a calculation of the $Q$  dependence of $G(\theta)$ for $\textbf{Q}$=(0.3 0.3 $Q_{l}$) for different anisotropic factors.  $G(\theta)$ is constant when the susceptibility is  isotropic ($\chi''_{a,b}=\chi''_{c}$) but shows different $Q$ dependence  with  anisotropy  ($\chi''_{a,b} \neq \chi''_{c}$). Namely,  $G(\theta)$ increases (decreases) with $Q$  when $\chi''_{c}/ \chi''_{a,b}$ $< 1 (>1)$. 

The $Q$ dependence of the total intensity $I$ given by Eq.(1) is shown for (0.3 0.3 $Q_{l}$), (0.7 0.3 $Q_{l}$) and (0.7 0.7 $Q_{l}$) in Fig. 1(c).  If $\chi''(\textbf{q},\omega)$ is isotropic,  the intensity $I$ is scaled only by $f_{Q}^2$ but it decreases slower (faster) than $f_{Q}^{2}$ in the presence of anisotropy  $\chi''_{c} / \chi''_{a,b}$ $< 1 (>1)$. This illustrates the fact that the anisotropic nature of the spin fluctuations can be directly determined by
the comparison of the $Q$ dependence of the intensity $I$ and  $f_{Q}^{2}$.
We would like to stress that knowing the magnetic form factor accurately is the key to the accurate determination of the spin susceptibility anisotropy factor.  Unfortunately the only magnetic form factor that can be found in the literature for the Ruthenium is that for Ru$^{+}$ ($f_{Q}$(Ru$^{+}$)) \cite{Ru-form}. The Ru ions in Sr$_{2}$RuO$_{4}$ are not Ru$^{+}$ but Ru$^{4+}$(nominally). Furthermore, Sr$_{2}$RuO$_{4}$ is not an insulator but an itinerant electron system, and the use of $f_{Q}$(Ru$^{+}$) to characterize the magnetic response of Sr$_{2}$RuO$_{4}$ is clearly inadequate. For this reason we decided to determine the magnetic form factor for Sr$_{2}$RuO$_{4}$ ($f_{Q}$(Sr$_2$RuO$_4$)) experimentally.

\subsection{Determination of the magnetic form factor of Sr$_{2}$RuO$_{4}$}

To determine the magnetic form factor of Sr$_{2}$RuO$_{4}$, $f_{Q}$(Sr$_{2}$RuO$_{4}$), we measured the $Q$ dependence of spin susceptibility at  several $\textbf{Q}$ positions with $\textbf{q}$$_{0}$=(0.3 0.3 0)  in the ($h$ $k$ 0) plane ($\theta$ = 0). The $Q$ dependence of the observed intensities is shown in Fig. 2, the filled and open symbols indicate our data and those reported data by Sidis \textit{et al.} \cite{Sidis99}, respectively. Note that, throughout the present paper all the quoted intensities have been corrected for resolution-volume effects, and that all the quoted experimental errors correspond to $2 \times \sigma$ in order to reflect the ambiguities of the scattering technique.  

Note that (0.7 0.3 0) and (1.3 0.3 0) are not $\textbf{q}_0$ positions from the reciprocal zone center, $\Gamma$ point, but those from the Z point (ex. (1 0 0)). These data can be treated equally with other data, because the spin susceptibility at  $\textbf{q}_0$ shows a strong two dimensionality and a rod type scattering along the $c^{*}$-axis  so that one can observe the signal even on the ($h$ $k$ 0) zone.

Our first observation of Fig. 2, is that it is clear that the data do not scale with $f_{Q}^{2}$(Ru$^{+}$), and decrease faster than it. This behavior is consistent with the fact that Sr$_{2}$RuO$_{4}$ is an itinerant electron system where mobile electrons distribute wider in real space than localized electron system and strongly indicates that one can not use $f_{Q}^{2}$(Ru$^{+}$) to evaluate anisotropic factor of spin susceptibility of Sr$_{2}$RuO$_{4}$. 

The full line in this figure is $f_{Q}^2$(Sr$_{2}$RuO$_{4}$) determined in the present study, for this purpose we fitted the observed intensities to the expression 
\begin{equation}
f_{Q}(\mathrm{Sr_{2}RuO_{4}})=A \ exp \ [B \ (Q/4 \pi)^{2}]+C.
\end{equation}
Here, we assumed that the $f_{Q}$ in the ($h$ $k$ 0) plane is isotropic, so that the $Q$ dependence of the $f_{Q}$ is described as a single $Q$ function \cite{Comment1}. 

This form factor was used to evaluate the anisotropic factor of the incommensurate spin fluctuations. Note that conductivity and coherence length of Sr$_2$RuO$_4$ show anisotropic behavior ($\sigma_{a,b} > \sigma_{c}$ and $\xi _{a,b}>\xi_{c} $) \cite{review2003}. Such results indicate that a distribution of unpaired electrons along the $c$-axis is confined and then the decrease of $f_{Q}$ with $Q_l$ must be slower than that for $Q_h$ or $Q_k$. It should be stressed here that we assumed an isotropic form factor $f_{Q}$ in all directions in the present study which causes an {\it underestimation} of the $c$-axis anisotropy.

\section{Experimental Results}

\subsection{$Q_l$ dependence of intensity}

In order to study the $Q_l$ ($\theta$) dependence of the intensity, we performed several series of constant-\textit{E} scans along  (0.3 0.3 $Q_l$), (0.7 0.7 $Q_l$), (0.7 0.3 $Q_l$) and (1.3 0.7 $Q_l$), and found that, because of low intensity,  it is difficult to get accurate $Q_l$ dependence at $\textbf{Q}$ positions  farther than (0.7 0.7 $Q_l$). For this reason we report only the results at the (0.3 0.3 $Q_l$)  and (0.7 0.3 $Q_l$) scans. Furthermore, to collect reliable data, one needs to select a clear window of energy where any spurious peaks including phonon scattering do not appear. For the constant-$E$ scans at (0.3 0.3 $Q_{l}$) and (0.7 0.3 $Q_{l}$), neutron transfer energies were selected to be 4 meV and 8 meV, respectively, by measurements of energy dependence of intensity at (0.3 0.3 0) and (0.7 0.3 0)  with  energy transfer between $\sim - 2 < E < \sim 20$ meV.  The energy dependence of intensity at (0.3 0.3 0) is shown in an inset of Fig. 3(a). The result clearly shows that the intensity at $E$ = 4 meV is affected by neither incoherent nor forward scattering. 

$Q_{l}$ dependence of integrated intensity at (0.3 0.3 $Q_{l}$) and (0.7 0.3 $Q_{l}$) are depicted in Fig. 3(a), the  integrated intensities were calculated as the  product of intensities at (0.3 0.3 $Q_{l}$), (0.7 0.3 $Q_{l}$) and the width determined by constant-$E$ scans along the ($h$ $k$ 0) direction. The obtained widths at ($h$ $h$ $Q_{l}$) and (0.7$h$ 0.3$h$ $Q_{l}$) were almost constant with $Q_l$, and we used their averaged values, cf. 0.188 and 0.184 \AA$^{-1}$ (in FWHM), respectively.  In addition, intensities at (0.5 0.5 $Q_{l}$) and (0.7$\pm$0.1 0.3$\pm0.04$ $Q_{l}$), which are almost constant with $Q_{l}$, were subtracted as  background for calculations of peak intensities at (0.3 0.3 $Q_{l}$) and (0.7 0.3 $Q_{l}$), respectively. Finally the data at (0.7 0.3 $Q_{l}$) with 8 meV were scaled with the data at (0.3 0.3 $Q_{l}$) with 4 meV by detailed measurements of energy dependence of signals.

The $Q_l$ dependence of the integrated intensity (0.3 0.3 $Q_{l}$) and (0.7 0.3 $Q_{l}$) in the Fig. 3(a) shows a very broad peak centered at $Q_{l}=0$ , indicating the strong two dimensionality of the spin fluctuations. This result is consistent with the one reported by Servant \textit{et al.} \cite{Servant01}, and allows us to neglect the  magnetic correlations along the $c$-axis. Thus we treat data sets  with different $Q_l$ independently.

\subsection{Determination of the anisotropic factor}

Fig. 3(b) shows the $Q$ dependence of the intensities for (0.3 0.3 $Q_{l}$) and (0.7 0.3 $Q_{l}$), the  full line is the magnetic form factor $f_{Q}^2$(Sr$_{2}$RuO$_{4}$) that we measured as indicated above. This figure clearly shows that the intensities for both (0.3 0.3 $Q_{l}$) and (0.7 0.3 $Q_{l}$) decrease faster  than $f_{Q}^{2}$(Sr$_{2}$RuO$_{4}$) with increasing $Q$. Such $Q$ dependence correspond to the case with $\chi''_{a,b}$ $< $ $\chi''_{c}$ as demonstrated in Fig. 1(c).

To evaluate the anisotropic factor, $Q$ dependence data for  (0.3 0.3 $Q_{l}$) was fitted to Eq. (1)(2), the data for (0.7 0.3 $Q_{l}$) was not used in this fit because of the poor statistics. From the fitting, we calculated the anisotropic factor of the spin susceptibility, $\chi''_{c}/\chi''_{a,b}  \sim 2.8 \pm 0.7$. Note that to evaluate an error in the determination of the anisotropic factor we took into account the error of the magnetic form factor. Furthermore, as explained in the previous section,  we assumed an isotropic form factor and such assumption may cause the  underestimation of the $c$-axis anisotropy. These results let us conclude that the incommensurate antiferromagnetic fluctuations observed at $\textbf{q}_0$ = (0.3 0.3 0) exhibit $c$-axis anisotropy. 

\section{Discussion}
\subsection{Static and dynamical spin susceptibility observed in Sr$_2$RuO$_4$}

The magnetic properties in the normal-state of Sr$_{2}$RuO$_{4}$ reported so far are 
(A) slightly anisotropic uniform susceptibility at $q$ = 0 \cite{Magnetization00},
(B) anisotropic spin fluctuations with anisotropic factor of $\chi''_{c}/\chi''_{a,b}  \sim 3$ at somewhere in $\textbf{q}$  reported by NMR \cite{IshidaYY},
(C) spin fluctuations observed at incommensurate $\textbf{q}$ of (0.3 \ 0.3 \ 0)  observed by INS \cite{Sidis99}.

The uniform susceptibility of (A) is explained by the Pauli paramagnetism of the conduction electrons in a two-dimensional $\gamma$ band, and the origin of a slight anisotropy of it ($\chi_{c}/\chi_{a,b}  \sim 1.1$) is attributed to the orbital Van Vleck contribution, which is affected by fields parallel to the $c$-axis due to the 1-dimensional $\alpha$- and $\beta$-bands \cite{KKNG00}. 

On the other hand, the anisotropy in (B) can not be associated with that in (A), because the anisotropic factor and energy scale of each spin susceptibilities are quite different. Judging from the similarity in  temperature dependence of spin fluctuations in (B) and (C), Ishida \textit{et al.} 
speculated that the anisotropic behavior observed in  the NMR measurements has a close relation with spin fluctuations observed at (0.3 0.3 0) \cite{IshidaYY}. Supporting this, our result clearly revealed that the incommensurate spin fluctuation has anisotropy with an anisotropic factor of $\chi''_{c}/\chi''_{a,b}  \sim 2.8$. The anisotropic factor reported by NMR measurement  is  $\sim 3$, which is in good agreement with the present result. These results  let us conclude that  anisotropic behavior observed by the NMR measurements is associated with spin fluctuations at  incommensurate $\textbf{q}$$_{0}$ vector of (0.3 0.3 0).

\subsection{The origin of the anisotropic behavior}

A short comment about the origin of the anisotropy of the incommensurate spin fluctuations observed at $\textbf{q}$ = (0.3 0.3 0). Theoretical calculations within the random-phase approximation using a three-band Hubbard Hamiltonian predict that  spin-orbit coupling plays an important role  and that, due to strong coupling, the out-of-plane component of the spin susceptibility is about two times larger than the in-plane one at low temperature \cite{Eremin00}. The calculated anisotropy and our result are quantitatively consistent. Magnetic property of (A), (B), and (C) deeply connect with the orbital of $d$-electrons in RuO$_{2}$ planes.  These facts strongly indicate that the spin-orbit interaction is important to discuss the magnetic properties of this system.

\subsection{Relation between incommensurate spin fluctuations and the superconducting mechanism}

As described in the introduction, some theoretical groups reported that incommensurate spin fluctuations with a $c$-axis anisotropy, $\chi''_{c}  > \chi''_{a,b}$, may introduce a spin triplet superconducting state and that the $d$ vector turns to a direction of larger antiferromagnetic fluctuations \cite{Kuwabara00,Sato00,Kuroki01}.
Our results show that the incommensurate spin fluctuations observed in Sr$_{2}$RuO$_{4}$ satisfy this requirement, namely $\chi''_{c} > \chi''_{ab}$, which makes  a direction of $d$ vector to be parallel to the $c$-axis consistent with the experimental observations \cite{Luke98,Ishida98,Duffy00}.

Then the question here is whether these spin fluctuations are really driving forces of the superconductivity of this material or not. Basically, the superconductivity of Sr$_{2}$RuO$_{4}$ is believed to originate in the quasi two-dimensional $\gamma $ main band. 
On the other hand, the incommensurate antiferromagnetic fluctuations of Sr$_{2}$RuO$_{4}$ is caused due to the nesting property of the one-dimensional $\alpha $ and $\beta $ bands. 
Furthermore, in the Sr$_{2}$Ru$_{1-x}$Ti$_{x}$O$_{4}$ (in which superconductivity is quickly suppressed and the antiferromagnetic fluctuations observed at $x$ = 0 develop into a static order with increasing $x$), the  $x$ dependence of $T_c$ seems to be explained only by a doping effect and no enhancement of $T_c$ by the spin fluctuations was  observed \cite{Kikugawa02,Braden02Ti,Ishida03}.  These results imply that the incommensurate spin fluctuations may not contribute to its superconducting mechanism \cite{Ishida03}. In order to further clarify the mechanism of the superconductivity in Sr$_{2}$RuO$_{4}$, especially of relations between spin triplet superconductivity and antiferromagnetic fluctuations, information of $\chi''(\textbf{q}_{0},\omega)$ behavior below $T_c$  would be of great help.

\subsection{Discrepancies with unpolarized INS results by other groups}

In the present study, we measured data at (0.3 0.3 $Q_l$) and (0.7 0.3 $Q_l$) including (0.3 0.3 0) ($Q \sim$ 0.70 \AA$^{-1}$) and estimated the anisotropic factor $\chi''_{c}/\chi''_{a,b}$ to be  $ \sim 2.8$ by  evaluating a difference between the $I$ and $f_{Q}^{2}$(Sr$_{2}$RuO$_{4}$). These results, however, are at odds with other unpolarized neutron scattering data reported by Servant \textit{et al.} \cite{Servant01} and by Braden \textit{et al.} \cite{Braden02}. We attribute these discrepancies to  (a) the narrower $Q$ range in these groups' measurements, and (b) a lack of a determination of  $f_{Q}^{2}$(Sr$_2$RuO$_{4}$) by the other groups.  
For example,  Servant $\textit{et al.}$ measured data at (0.3 0.3 $Q_l$) with only larger $Q$ part ($Q > 0.80$ \AA$^{-1}$) and (0.7 0.7 $Q_l$), and  concluded an isotropic behavior of spin fluctuations based on the fact that a small number of data points observed along (0.3 0.3 $Q_l$) and (0.7 0.7 $Q_l$) scaled at a very narrow $Q$ range near $Q$ of $\sim$ 1.6 \AA$^{-1}$. One can see in Fig. 3(b) of our paper, that the accuracy of the data in that $Q$ range (near $\sim$ 1.6 \AA$^{-1}$) is not very good.  We also observed data at (0.7 0.7 $Q_l$)  and found that the data scaled with those along (0.3 0.3 $Q_l$) in this $Q$ range within the huge error bars. Furthermore, they did not get the proper magnetic form factor for Sr$_2$RuO$_{4}$, this prevented them from making a reliable comparison of their data with the magnetic form factor in the small $Q$ region. 
On the other hand, Braden \textit{et al.} observed $Q$-dependence of $I$ at (0.3 0.3 $Q_l$) within a very narrow $Q$ range of 1.2 $\sim$ 2.5 \AA$^{-1}$ and showed that the data decreases slower than that of $f_{Q}^{2}$(Ru$^+$) \cite{Braden02}. This behavior is clearly opposite to the our data shown in this paper and to the data of Servant \textit{et al.} \cite{Servant01}. At this time we do not understand the source of this discrepancy.

Recently, neutron polarization analysis experiments have been performed on Sr$_2$RuO$_{4}$ by two independent groups. These groups succeeded in confirming the $c$-axis anisotropy with anisotropic factor of 2$\sim$2.5 \cite{cm0307662} and 2.0 $\pm$ 0.4 \cite{cm0308558}), respectively, being consistent with our unpolarized neutron results presented here.
It can be argued that the best way to perform this type of measurements is using the polarized neutron scattering technique because this technique allows the separation of  magnetic components to the scattering from any other non-magnetic components including phonon and spurious peaks. We would like to stress here, however, that this is not the only reliable way to measure magnetic components. It is true that the unpolarized neutron technique intrinsically more ambiguous when it comes to measure magnetic components. But being conscious of this fact, we paid the greatest care to reduce such errors and we made many consistency checks with different scattering zones and even checked background from the cryostat and judiciously chose the best conditions for the experiment. Every experimental result presented in our paper has been examined with great caution and our results are reliable.

\section{Summary}

We have performed unpolarized inelastic neutron scattering measurements on Sr$_2$RuO$_{4}$ to probe the anisotropic behavior of the spin susceptibility observed at the incommensurate wave vector of $\textbf{q}$ = (0.3 0.3 0). Our measurements indeed support that the susceptibility exhibits a $c$-axis anisotropy i.e. $\chi''_{c}/\chi''_{a,b} \sim 2.8 \pm 0.7$.
This anisotropy ratio is in good agreement with  the result obtained by the NMR measurements ($\sim 3$) \cite{IshidaYY}.

\begin{acknowledgments}

We would like to acknowledge Dr. J. A. Fernandez-Baca for valuable discussions and for a 
critical reading of the manuscript. T. N., H. F. and H. Y. were supported by a Grant-In-Aid from the 
Ministry of Education, Culture, Sports, Science and Technology, Japan.

\end{acknowledgments}

\begin{figure}[htpd]
\includegraphics[width=6cm]{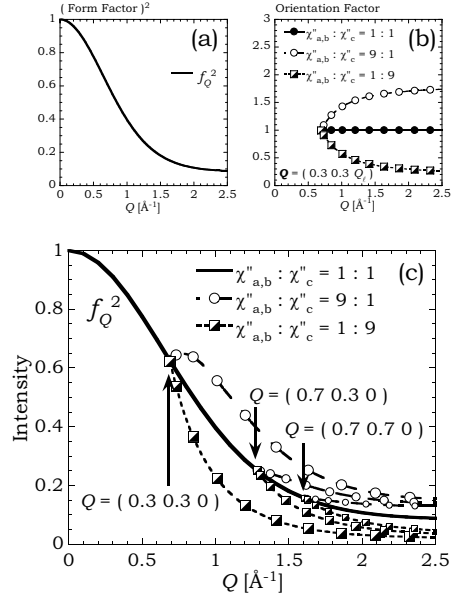}
\caption{\label{fig030724-1}
(a) A schematical $Q$ dependence of the square of magnetic form factor,  
$f^2_{Q}$. (b) $Q$ dependence of calculated orientation factors  
$G(\theta)$ at $\textbf{Q}$ = (0.3 0.3 $Q_{l}$) for different anisotropic factors. Depending on a ratio, $\chi''_{a,b}$ : $\chi''_{c}$, $G(\theta)$ shows different $Q$ dependence.
(c) $Q$ dependence of calculated intensities $I$  ($= f_{Q}^{2} \times  G(\theta)$) at (0.3 0.3 $Q_{l}$), (0.7 0.3 $Q_{l}$) and (0.7 0.7 $Q_{l}$) with different anisotropic factors.
}
\end{figure}
\begin{figure}[htpd]
\includegraphics[width=6cm]{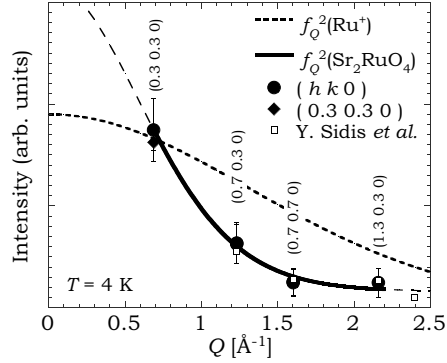}
\caption{\label{fig030724-2}
$Q$ dependence of intensities observed at $\textbf{Q}$  with $\textbf{q}_0$=(0.3 0.3 0) in the  
($h$ $k$ 0) plane ($\theta$ = 0). Filled symbols are the present results, 
in which circle and diamond symbols correspond to the data taken with 
different sample set with ($h$ $k$ 0) and ($h$ $h$ $l$) with $l$ = 0,  
respectively. Open symbols are taken from Ref. \cite{Sidis99} reported by Sidis  
\textit{et al.}. Taking into accounts ambiguities of scattering experiments, 
we conservatively adopt $2 \times \sigma$ error bars.  
A broken  and a full lines correspond to the square of  
magnetic form factor of Ru$^{+}$, $f^2_{Q}$(Ru$^{+}$), and that of  
Sr$_{2}$RuO$_{4}$, $f_{Q}^{2}$(Sr$_{2}$RuO$_{4}$), respectively.  
The latter was determined and parameterized with Eq. (3) in the present study (See the main text).}
\end{figure}
\begin{figure}[htpd]
\includegraphics[width=6cm]{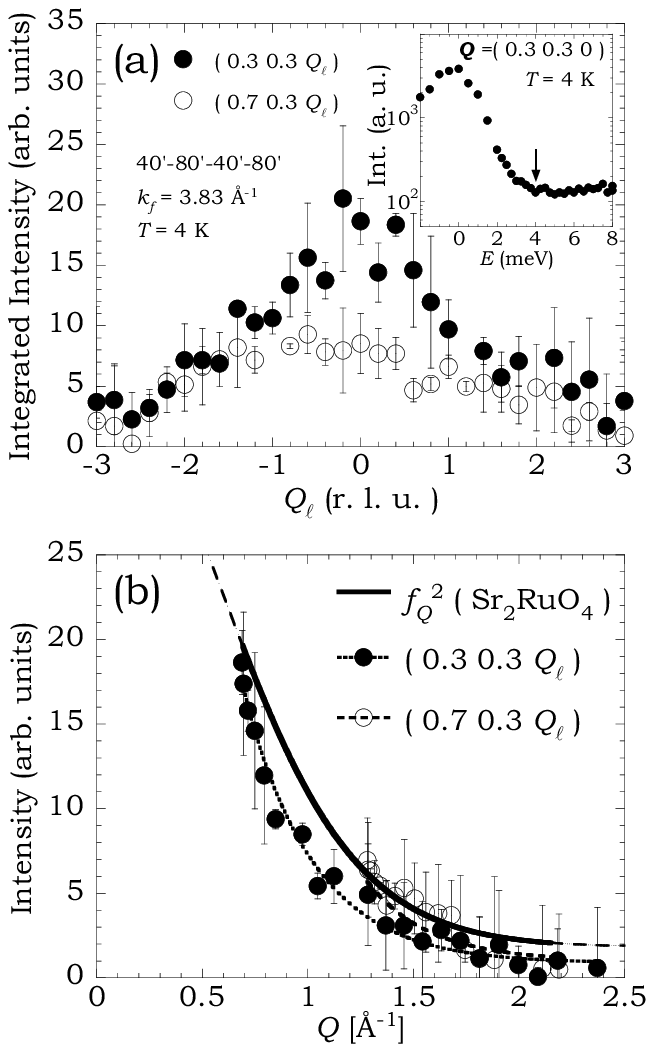}
\caption{\label{fig030724-3}
(a) $Q_{l}$ dependence of integrated intensity $I$ at (0.3 0.3 $Q_{l}$) 
and (0.7 0.3 $Q_{l}$) and at 4 K. Some data points are missing because of uncertainty 
due to existence of spurious peaks. (inset) : Energy dependence of scattering  
intensity at (0.3 0.3 0). 
(b) $Q$ dependence of resolution corrected intensity at (0.3 0.3 $Q_{l}$) and (0.7 0.3 $Q_{l}$). 
Averaged values  between ($Q_h$ $Q_k$  $|Q_{l}|$) and ($Q_h$ $Q_k$ $-|Q_{l}|$) are plotted. 
The full line is the square of magnetic form factor of  Sr$_{2}$RuO$_{4}$,  
$f_{Q}^{2}$(Sr$_{2}$RuO$_{4}$), determined in the present study (See  
Fig. 2). Dashed line on (0.3 0.3 $Q_{l}$) is a fitting curve to Eq. (1)(2) with  
$\chi''_{c}/\chi''_{a,b} = 2.8$ and a broken line on (0.7 0.3 $Q_{l}$) is a simulation 
line calculated with the parameters evaluated by data at (0.3 0.3 $Q_{l}$). In both plots,
 $2 \times \sigma$ was conservatively adopted as error bars.
}
\end{figure}


\begin{references}

\bibitem{Maeno94} Y. Maeno, H. Hashimoto, K. Yoshida, S. Nishizaki, T. Fujita, J. G. Bednorz, and F. Lichtenberg, Nature (London) \textbf{372}, 532 (1994).

\bibitem{review2003} A. P. Mackenzie and Y. Maeno, Rev. Mod. Phys. \textbf{75}, 657 (2003).

\bibitem{Luke98} G. M. Luke, Y. Fudamoto, K. M. Kojima, M. I. Larkin, J. Merrin, B. Nachumi, Y. J. Uemura, Y. Maeno, Z. Q. Mao, Y. Mori, H. Nakamura, and M. Sigrist, Nature (London) \textbf{394}, 558 (1998).

\bibitem{Ishida98} K. Ishida, H. Mukuda, Y. Kitaoka, K. Asayama, Z. Q. Mao, Y. Mori, and Y. Maeno, Nature (London) \textbf{396}, 658 (1998).

\bibitem{nonmagimpurity}  A. P. Mackenzie, R. K. W. Haselwimmer, A. W. Tyler, G. G. Lonzarich, Y. Mori, S. Nishizaki, and Y. Maeno,  Phys. Rev. Lett. \textbf{80} 161 (1998).



\bibitem{Sidis99} Y. Sidis, M. Braden, P. Bourges, B. Hennion, S. NishiZaki, Y. Maeno, and Y. Mori, Phys. Rev. Lett. \textbf{83}, 3320 (1999).

\bibitem{Servant01} F. Servant, B. F\r{a}k, S. Raymond, J. P. Brison, P. Lejay, and J. Flouquet,  Phys. Rev. B \textbf{65}, 184511 (2002).

\bibitem{Braden02} M. Braden, Y. Sidis, P. Bourges, P. Pfeuty, J. Kulda, Z. Mao, and Y. Maeno, Phys. Rev. B \textbf{66}, 064522 (2002).

\bibitem{Upt3} H. Tou, Y. Kitaoka, K. Ishida, K. Asayama, N. Kimura, Y. Onuki, E. Yamamoto, Y. Haga, and K. Maezawa,  Phys. Rev. Lett. \textbf{80}, 3129 (1998).

\bibitem{Uni2Al3} K. Ishida, D. Ozaki, T. Kamatsuka, H. Tou, M. Kyogaku, Y. Kitaoka, N. Tateiwa, N. K. Sato, N. Aso, C. Geibel, and F. Steglich, Phys. Rev. Lett. \textbf{89}, 037002 (2002).








\bibitem{Mazin97} I. I. Mazin and D. J. Singh, Phys. Rev. Lett. \textbf{79}, 733 (1997).

\bibitem{Mackenzie96} A. P. Mackenzie, S. R. Julian, A. J. Diver, G. J. McMullan, M. P. Ray, G. G. Lonzarich, Y. Maeno, S. Nishizaki, and T. Fujita, Phys. Rev. Lett. \textbf{76}, 3786 (1996).

\bibitem{Mazin99} I. I. Mazin and D. J. Singh, Phys. Rev. Lett. \textbf{82}, 4324 (1999).


\bibitem{Kuwabara00} T. Kuwabara and M. Ogata, Phys. Rev. Lett. \textbf{85}, 4586 (2000).
\bibitem{Sato00} M. Sato and M. Kohmoto,  J. Phys. Soc. Jpn. \textbf{69}, 3505 (2000).
\bibitem{Kuroki01} K. Kuroki, M. Ogata, R. Arita, and H. Aoki, Phys. Rev. B \textbf{63}, 060506 (2001).


\bibitem{IshidaYY} K. Ishida, H. Mukuda, Y. Minami, Y. Kitaoka, Z. Q. Mao, H. Fukazawa, and Y. Maeno, Phys. Rev. B \textbf{64}, 100501 (2001).





\bibitem{comments02} In order to keep effective sample volume constant, we tried to assemble single crystals symmetrically. Then, to check a reliability, we have measured a ratio of intensities between (0.3 0.3 0) and (0.7 0.7 0) with different sample sets with different scattering planes ($h$ $k$ 0) and ($h$ $h$ $l$), and found that  the ratio is same within experimental accuracy. We also checked absorption factors of Sr, Ru and O atoms in a text book. They are negligibly small. From these facts, we concluded that geometrical corrections are not necessary in the present experiments, and we just corrected the observed intensity by the instrumental resolution.  

\bibitem{Book} As a text, see Neutron Scattering with a Triple-Axis Spectrometer (Basic Techniques), G. Shirane, S. M. Shapiro and J. M. Tranquada, Cambridge university press.




\bibitem{Ru-form} $International\ Tables\  of\  Crystallography$, edited by A. J. C. Wilson (Kluwer Academic, Dordrecht, 1995), Vol. C.

\bibitem{Comment1} If there is a remarkable anisotropy in the in-plane susceptibility, data at (0.3 0.3 0) ($Q$ = 0.69 \AA$^{-1}$) and (0.7 0.7 0) ($Q$ = 1.61 \AA$^{-1}$) and that of (0.7 0.3 0) ($Q$ = 1.25 \AA$^{-1}$) draw different lines. 
Within experimental accuracy, such a distinct behavior is not observed,  indicating that the sizable anisotropy of magnetic form factor in the RuO$_{2}$ plane was not detected. 

\bibitem{Magnetization00} Y. Maeno, K. Yoshida, H. Hashimoto, S. Nishizaki, S. Ikeda, M. Nohara, T. Fujita, A. P. Mackenzie, N. E. Hussey, J. G. Bednorz, and F. Lichtenberg, J. Phys. Soc. Jpn. \textbf{66}, 1405 (1997).


\bibitem{KKNG00} K. K. Ng and M. Sigrist, J. Phys. Soc. Jpn. \textbf{69}, 3764 (2000).

\bibitem{Eremin00} I. Eremin, D. Manske, and K. H. Bennemann, Phys. Rev. B \textbf{65} 220502 (2002).


\bibitem{Duffy00} J. A. Duffy, S. M. Hayden, Y. Maeno, Z. Mao, J. Kulda, and G. J. McIntyre, Phys. Rev. Lett. \textbf{85}, 5412 (2000).


\bibitem{Kikugawa02} N. Kikugawa and Y. Maeno, Phys. Rev. Lett. \textbf{89}, 117001 (2002).

\bibitem{Braden02Ti} M. Braden, O. Friedt, Y. Sidis, P. Bourges, M. Minakata, and Y. Maeno, Phys. Rev. Lett. \textbf{88}, 197002 (2002).

\bibitem{Ishida03}  K. Ishida, Y. Minami, Y. Kitaoka, S. Nakatsuji, N. Kikugawa, and Y. Maeno, Phys. Rev. B \textbf{67} 214412 (2003).


\bibitem{cm0307662} M. Braden, P. Steffens, Y. Sidis, J. Kulda, S. Hayden, N. Kikugawa, and Y. Maeno, cond-mat/0307662.

\bibitem{cm0308558} B. F\r{a}k, S. Raymond, F. Servant, P. Lejay, and J. Flouquet, cond-mat/0308558.





\end{references}
\end{document}